\documentclass[reqno]{amsart}
\numberwithin{equation}{section}
 
 \DeclareMathOperator{\diag}{\rm diag}
 \DeclareMathOperator{\Cs}{\mathbb{C}}
 \DeclareMathOperator{\Ns}{\mathbb{N}}
 \DeclareMathOperator{\Rs}{\mathbb{R}}
 \DeclareMathOperator{\Zs}{\mathbb{Z}}
 
 \DeclareMathOperator{\Co}{\mathcal{C}}

 \DeclareMathOperator{\Io}{\mathcal{I}}
 \DeclareMathOperator{\Lo}{\mathcal{L}}
 \DeclareMathOperator{\Mo}{\mathcal{M}}
 
 \DeclareMathOperator{\Ss}{\mathcal{S}}
 \DeclareMathOperator{\To}{\mathcal{T}}
 \DeclareMathOperator{\Uo}{\mathcal{U}}
 \DeclareMathOperator{\q}{\mathbf{q}}
 
 \DeclareMathOperator{\bz}{\mathbf{z}}

\begin{document}
\title[]{2D Toda chain and associated commutator
identity}
\author[]{A.~K.~Pogrebkov}
\address[A.~K.~Pogrebkov]{Steklov Mathematical Institute, Moscow, Russia}
\email{pogreb@mi.ras.ru}
\thanks{This work is supported in part by the Russian Foundation for Basic Research (grant \# 05-01-00498 and grant
\# 06-01-92057-CE), Scientific Schools 672.2006.1, by NWO--RFBR (grant \# 047.011.2004.059), by the Program
of RAS ``Mathematical Methods of the Nonlinear Dynamics''.}
\date{\today}

\begin{abstract}
Developing observation made in~\cite{commut} we show that simple identity of the commutator type on an
associative algebra is in one-to-one correspondence to 2D (infinite) Toda chain. We introduce representation
of elements of associative algebra that, under some generic conditions, enables derivation of the Toda chain
equation and its Lax pair from the given commutator identity.
\end{abstract}
\maketitle

\section{Introduction. Commutator identity and linearized version of the Toda chain}\label{intro}

In~\cite{commut} we demonstrated that well know integrable nonlinear pde's, such as KP, BLP, DS,
Sine-Gordon, Nonlinear Schr\"odinger are in mutual correspondence with some simple commutator identities on
associative algebras. Say, identity associated to the Kadomtsev--Petviashvili equation has the form
\begin{equation}
 [A_{}^{3},[A,B]]-\dfrac{3}{4}[A_{}^{2},[A_{}^{2},B]]-\dfrac{1}{4}[A,[A,[A,[A,B]]]]=0.\label{commut1}
\end{equation}
We also suggested there representation of elements $A$ and $B$ as extended operators in terms of the
approach developed in~\cite{first}--\cite{KP-JMP}, that can be considered as a version of the dressing
procedure. Then we proved that under some general conditions imposed on this representation one can uniquely
reconstruct both, nonlinear equations themselves and their Lax pairs, once the commutator identities are
given. Here we generalize this construction for lattice models using 2D Toda chain as example.

As well as in case of~(\ref{commut1}) it is easy to check that any two arbitrary elements $A$ and $B$ of an
arbitrary associative algebra under assumption of invertability of $A$ obey identity
\begin{equation}\label{1}
  [A,[A^{-1},B]]=2B-ABA^{-1}-A^{-1}BA.
\end{equation}
This identity, strictly speaking, is not a commutator one (in contrast to~(\ref{commut1})), as the r.h.s.\
involves similarity transformations of $B$. But substituting here $A\to e^{\alpha A}$ and differentiating 4
times with respect to $\alpha$ at zero we get~(\ref{commut1}). For this reason we refer to~(\ref{1}) as
commutator identity as well. Like identities considered in~\cite{commut} this identity appears as the lowest
term of the (semi)infinite hierarchy
\begin{equation}\label{2}
 [A_{}^{-n},\underbrace{[A,\ldots,[A,}_{n}B]\ldots]=
 \sum_{m=0}^{n}\dfrac{n!(-1)^{m-n}}{m!(n-m)!}A^{m}(A^{-n}BA^{n}-B)A^{-m},\quad n=1,2,\ldots.
\end{equation}
Existence of such identity proves that an infinite set of functions of $t=\{t_1,t_2\}$ defined as
\begin{equation}\label{3}
  B(m,t_1,t_2)=e^{-t_1A+t_2A^{-1}}A^{m}BA^{-m}e^{t_1A-t_2A^{-1}},\quad m=0,\pm1,\pm2,\ldots,
\end{equation}
obeys differential equations
\begin{equation}\label{4}
  \dfrac{\partial^{2}B(m,t)}{\partial{t_1}\partial{t_2}}=B(m+1,t)+B(m-1,t)-2B(m,t).
\end{equation}
This system is a linearized version of the famous integrable system in (2+1)-dimesions (where one of
variables is discrete): 2D infinite Toda chain,
\begin{equation}\label{5}
  \dfrac{\partial^{2}\phi_{n}}{\partial{t}\partial{x}}=e^{\phi_{n+1}-\phi_{n}}_{}-e^{\phi_{n}-\phi_{n-1}}_{},\quad n\in\Zs,
\end{equation}
that is known to have Lax pair
\begin{equation}\label{6}
  \partial_{t}\Lo=[\Lo,\Mo],
\end{equation}
where infinite matrix differential operators $\Lo$ and $\Mo$ equal
\begin{align}\label{7}
  \Lo&= \partial_{x}\Io+\To-\mathcal{U}, \\
  \Mo&=\To^{-1}\mathcal{C}.\label{8}
\end{align}
Here $\Io$ and $\To$ are unity and shift matrices,
\begin{equation}\label{9}
  \Io_{m,n}=\delta_{m,n},\qquad\To_{m,n}=\delta_{m-1,n},\quad m,n\in\Zs,
\end{equation}
and $\mathcal{U}$ and $\mathcal{C}$ are two diagonal matrices
\begin{align}
 & \Uo=\diag\{\partial_{x}\phi_{n}\},\label{10:12} \\
 & \Co=\diag\{e^{\phi_{n}-\phi_{n-1}}\},\label{10:13}
\end{align}
that obey
\begin{align}
 & \Uo_{t}=\To^{-1}\Co\To-\mathcal{C},\label{16:1} \\
 & \Co_{x}=\Co(\To\Uo\To^{-1}-\Uo).\label{16:2}
\end{align}
These relations follow from~(\ref{4}), (\ref{10:12}), and~(\ref{10:13}) and also can be considered as
another form of the Toda lattice equations (strictly speaking, more general than~(\ref{4})).

Under substitution~(\ref{2}) commutators of $A$ and $B$ in identity~(\ref{1}) correspond to differentiations
in~(\ref{3}), while combinations $B-ABA^{-1}$ and $B-A^{-1}BA$ are responsible for difference parts.
Defining analogously to~(\ref{3})
\begin{equation}\label{11}
  B(m,t_1,t_2,t_3,\ldots)=e^{-t_1A+t_2A^{-1}+t_3A^{-2}+\ldots}A^{m}BA^{-m}e^{t_1A-t_2A^{-1}-t_3A^{-2}-\ldots},
\end{equation}
we get that these functions thanks to~(\ref{2}) obey linearized versions of the highest equations of the
Toda chain hierarchy:
\begin{equation}\label{12}
 \dfrac{\partial^{n+1}B(k,t)}{\partial^{n}_{t_1}\partial_{t_n}}=
 \sum_{m=0}^{n}\dfrac{n!(-1)^{m}}{m!(n-m)!}(B(k+m-n,t)-B(k+m,t)),\quad n=1,2,\ldots,
\end{equation}
for any $k\in\Zs$.

Here our aim is to develop a kind of dressing procedure that enables reconstruction of the Toda chain
itself, as well as its Lax pair by means of the commutator identity~(\ref{1}). To solve this problem we need
to choose a special representation of elements $A$ and $B$. Taking into account that thanks to~(\ref{3})
\begin{align}
 & B(m+1,t)=AB(m,t)A^{-1},\label{13:1}\\
 &\partial_{t_1}B(m,t)=-[A,B(m,t)],\qquad\label{13:2}\\
 &\partial_{t_2}B(m,t)=[A^{-1},B(m,t)],\label{13:3}
\end{align}
we see that it is reasonable to realize this elements as infinite matrix integral operators. In Sec.~2 we
introduce mathematical details of this construction based on the method of the extended resolvent,
see~\cite{first}--\cite{KP-JMP}. In Sec.~3 we perform this realization and impose conditions that enables
uniqueness of this procedure. In Sec.~4 we introduce dressing transformation and derive Lax pair and
Eq.~(\ref{4}) by means of our construction. Concluding remarks are given in Sec.~5.

\section{Space of operators}\label{resolvent}
We work in the space of infinite matrix integral operators $F(q,h)$, $G(q,h)$, etc, with kernels,
correspondingly, $F_{m,m'}(x,x';q,h)$ and $G_{m,m'}(x,x';q,h)$, where $x,x',q,h\in\Rs$ and $h>0$,
$m,m'\in\Zs$. We assume that all these kernels belong to the space $\Ss'$ of distributions with respect to
all their variables. This enables introduction of the ``shifted'' Fourier transform,
\begin{align}
 F(p,\zeta;\q,\bz)&=\frac{1}{2\pi}\sum_{m,m'}e_{}^{i(m-m')\arg\bz}\zeta^{m}\int dx\int dx'\times\nonumber\\
 &\times e_{}^{i(p+{\q}_{\Re})x-i{\q}_{\Re}x'}F_{m,m'}(x,x';{\q}_{\Im},|\bz|), \label{2a} \\
\intertext{where $p\in\Rs$, $\q=\q_{\Re}+i\q_{\Im}\in\Cs$, $\zeta,\zeta',\bz\in\Cs$ and
$|\zeta|=|\zeta'|=1$. The inverse transformation is given by}
 F_{m,m'}(x,x';\q_{\Im},|\bz|) &=\frac{1}{2\pi}\int dp\int d{\q}_{\Re}\oint\limits_{|\zeta|=1}
 \frac{d\zeta\,\zeta_{}^{-m-1}}{2\pi i}
 \oint\limits_{|\zeta'|=1}\frac{d\zeta'\,{\zeta'}^{m'-1}} {2\pi i}\times\nonumber\\
 &\times e_{}^{-i(p+{\q}_{\Re})x+i{\q}_{\Re}x'} F(p,\zeta/\zeta';\q,|\bz|\zeta'),\label{2b}
\end{align}
Variables $q=\q_{\Im}$ and $h=|\bz|$ play role of parameters of operators and they are not involved in thier
composition:
\begin{align}
 &(FG)_{m,m'}(x,x';q,h)=\sum_{n}\int dy\,F_{m,n}(x,y;q,h)G_{n,m'}(y,x';q,h),\label{comp}\\
\intertext{that is defined for such pares of operators where integral in the r.h.s.\ exists in the sense of
distributions. In the $(p;\zeta)$-space this convolution takes the form}
 &(FG)(p;\q)=\int dp'\oint\limits_{|\zeta'|=1}\frac{d\zeta'}{2\pi i\zeta'}\,
 F(p-p',\zeta/\zeta';\q+p',\bz\zeta')G(p',\zeta';\q,\bz).\label{180}
\end{align}

Differential--difference operators are imbedded in this space of operators as a special subclass by means of
the following procedure. To any given infinite matrix differential operator $\Lo_{m,m'}(x,\partial_{x})$,
where $\Lo_{m,m'}(x,\partial_{x})$ equals to zero when difference $m-m'$ is large enough, with kernel
\begin{equation}
 \Lo_{m,m'}(x,x')=\Lo_{m,m'}(x,\partial_{x})\delta (x_{1}-x_{1}')\ldots\delta(x_{d}-x_{d'}')  \label{kern}
\end{equation}
we associate the \textbf{extended} operator with the kernel
\begin{equation}
  L_{m,m'}(x,x';q,h)=e_{}^{-q(x-x')}h_{}^{m-m'}\Lo_{m,m'}(x,x').  \label{1'}
\end{equation}
Then the kernel $L(p;\q,\bz)$ of this differential--difference operator (as given in~(\ref{2a})) depends on
variables $\q$, $\bz$ and $1/\bz$ polynomially. Thanks to~(\ref{2a}) for the ($p,\zeta$)-kernel of the
extension of a differential--difference operator $\Lo_{m,m'}$ we have
\begin{equation}\label{2aa}
  L(p,\zeta;\q,\bz)=\frac{1}{(2\pi)^{d}}\sum_{m,m'}\bz_{}^{m-m'}\zeta^{m}\int dx\int dx'
  e_{}^{i(p+\q)x-i\q x'}\Lo_{m,m'}(x,x').
\end{equation}

Consider some examples of such extended operators. Let $D(q,h)$ denote the extension of the differential
operator $i\partial_{x}$, i.e.,
\begin{align}
 &D_{m,m'}(x,x';q,h)=i(\partial_{x}+q)\delta (x-x')\delta_{m,m'},\label{10}\\
\intertext{while in $(p,\zeta)$-space kernel of this operator equals}
 &D(p;\q)=\q\delta (p)\delta_{c}(\zeta).  \label{10a}
\end{align}
Here we introduced the $\delta$-function on the contour $|\zeta|=1$,
\begin{equation}
 \delta _{c}^{}(\zeta)=\sum_{n=-\infty }^{\infty }\zeta_{}^{n},\label{18}
\end{equation}
so that we have
\begin{equation}
 \oint\limits_{|\zeta|=1}\frac{d\zeta}{2\pi i\zeta}\delta _{c}^{}(\zeta)f(\zeta)=f(1)  \label{19}
\end{equation}
for an arbitrary test function  $f(\zeta)=\sum\zeta^{m}f_{m}$ on the contour.

Correspondingly, operator $I$ that is the unity with respect to the composition~(\ref{comp}), (\ref{180})
has kernels
\begin{equation}\label{unity}
  I_{m,m'}(x,x';q,h)=\delta(x-x')\delta_{m,m'},\qquad I(p,\zeta;\q,\bz)=\delta(p)\delta_{c}(\zeta).
\end{equation}
Next, let $\To$ be the shift matrix as defined in~(\ref{9}), then by~(\ref{kern}), (\ref{1'}), (\ref{2aa})
and~(\ref{18}) its extension is given by kernels
\begin{equation}\label{T6}
  T_{m,n}(x,x';q,h)=h\delta_{m-1,n}\delta(x-x'),\qquad T(p,\zeta;\q,\bz)=\bz\delta(p)\delta_{c}(\zeta).
\end{equation}
The analyticity properties of kernels of differential--difference operators motivate introduction of
$\overline\partial$-differentiation with respect to complex variables $\q$ and $\bz$. For any given operator
$F$ with kernel $F(p,\zeta;{\q},\bz)$ we define two new operators, $\overline{\partial}_{\q}F$ and
$\overline{\partial}_{\bz}F$ given by their kernels
\begin{align}
 &(\overline{\partial}_{\q}F)(p,\zeta;{\q},\bz)=\dfrac{\partial F(p,\zeta;{\q},\bz)}
 {\partial\overline{\q}},  \label{2a12}\\
 &(\overline{\partial}_{\bz}F)(p,\zeta;{\q},\bz)=\dfrac{\partial F(p,\zeta;{\q},\bz)}
 {\partial\overline{\bz}},  \label{2a13}
\end{align}
where derivatives are understood in the sense of distributions. Then characterization property of the
differential--difference operators is given by conditions
\begin{equation}\label{dbL}
  \overline{\partial}_{\q}^{}F=0,\qquad\overline{\partial}_{\bz}^{}F\sim\delta(\bz)\text{ and its derivatives},
\end{equation}
where $\delta$-function of complex variable $\bz$ is defined by the standard equality
$\delta(\bz)=\delta(\bz_{\Re})\delta(\bz_{\Im})$. Let us mention that if $F$ is a differential--difference
operator with constant coefficients then by~(\ref{1'}):
\begin{equation}\label{L01}
  F(p,\zeta;\q,\bz)=f(\q,\bz)\delta(p)\delta_{c}(\zeta),
\end{equation}
where $f(\q,\bz)$ is polynomial function of $\q$, $\bz$, and $1/\bz$. Then by~(\ref{180}) we get for
commutator with arbitrary operator $F$ of considered class
\begin{equation}\label{L0F}
  [F,G](p,\zeta;\q,\bz)=(f(p+\q,\bz\zeta)-f(\q,\bz))G(p,\zeta;\q,\bz).
\end{equation}

\section{Operator representation of elements of associative algebra}\label{2+1}
We realize elements of associative algebra as operators $A_{m,m'}(q,h)$, $B_{m,m'}(q,h)$, etc., in the sense
of definitions of Sec.~\ref{resolvent} and we impose specific conditions on this realization that enables
derivation of the Lax pairs and nonlinear integrable equation. Identities~(\ref{1}) and~(\ref{2}) show that
commutator and similarity transformation with $A$ are generating elements of these relation in the sense
that commutator with $A^{-n}$ is given in their terms. Thus, following a generic procedure suggested
in~\cite{commut} we impose conditions:

\textbf{Condition 1.} Switching on of time dependence by means of one of relation~(\ref{3}) (or~(\ref{11}))
gives operator $B_{n,n'}(t,m,q,h)$ with kernels $B_{n,n'}(x,x';t,m,q,h)$ belonging to the same space of
operators.

Next, taking into account that in~(\ref{3}) variables corresponding to these two generators were denoted as
$t_1$ and $m$, we impose

\textbf{Condition 2.} The dependence of $B(m,t,q,h)$ with respect to variables $m$ and $t_1$ reduces to
the shift of the $x$ and matrix variables of the kernel, i.e.,
\begin{equation}\label{cond:tx}
  B_{n,n'}(x,x';m,t_1,t_2,q,h) =B_{n+m,n'+m}(x+t_1,x'+t_1,t_2;q,h).
\end{equation}

In terms of $(p,\zeta)$-kernels defined by~(\ref{2a}) this means that
\begin{align}
 &B(p,\zeta;m,t_1,t_2,\q,\bz)=e^{-it_1p}\zeta^{m}B(p,\zeta;t_2,\q,\bz),\label{cond:tp}\\
\intertext{or in the operator form}
 \label{Bn1}
  &B(m+1,t_1,t_2,q,h)=TB(m,t_1,t_2,q,h)T^{-1},\\
  &\partial_{t_{1}}B(m,t_1,t_2,q,h)=-i[D,B(m,t_1,t_2,q,h)],\label{Bn2}
\end{align}
where~(\ref{10a}) and~(\ref{T6}) were used. Finally, we impose

\textbf{Condition 3.} If dependence of the kernel $A(p,\zeta;\q,\bz)$ on $\q$ and $\bz$ variables reduces to
the only one linear combination then the same is dependence of the kernel $B(p,\zeta;\q,\bz)$. Formally this
condition means that operator $B$ is function of space variables and operator $A$.

We consider  now consequences of these conditions for operator representation of~(\ref{1}) and~(\ref{3}).
First, Eq.~(\ref{Bn1}) and~(\ref{13:2}) show that we can choose
\begin{equation}\label{AT}
  A=T,
\end{equation}
that, in fact, is unique choice if we take into account that thanks to~(\ref{T6}) operator $T$ is extension
of the shift matrix $\To$ in~(\ref{9}). This means that it includes additional multiplier $h$ that cancels
out in~(\ref{Bn1}).

By~(\ref{13:2}) and~(\ref{Bn2}) we have to obey equality
\[[T,B(m,t_1,t_2,q,h)]=-i[D,B(m,t_1,t_2,q,h)]\],
where thanks to~(\ref{3}) it is enough to put  $m$, $t_1$ and $t_2$ equal to zero. This means that there
exists operator
\begin{equation}\label{L0}
  L_{0}=-iD+T,
\end{equation}
such that
\begin{equation}\label{L0B}
  [L_0,B]=0.
\end{equation}
Because of~(\ref{10}) and~(\ref{T6}) operator $L_0$ is extension in the sense of~(\ref{2aa}) of the matrix differential operator
\begin{equation}\label{Lo0}
  \Lo_0=\Io\partial_{x}+\To,
\end{equation}
that is the special case of the Lax operator~(\ref{7}) of the Toda chain corresponding to potential $\Uo\equiv0$.

Next, thanks~(\ref{T6}), (\ref{AT}) the kernel of $A(p,\zeta;\q,\bz)$ is independent of variable $\bz$. Then
by Condition~3 the kernel of operator $B$ is also independent of $\bz$. By~(\ref{L0F}) in
terms of the kernels equality~(\ref{L0B}) sounds as
\begin{equation}\label{L0B'}
  (\bz(\zeta-1)-ip)B(p,\zeta;\bz)=0,
\end{equation}
i.e., kernel of operator $B$ has representation
\begin{equation}\label{toda29:3}
    B(p,\zeta;\q,\bz)=\delta\!\left(\bz-\dfrac{ip}{\zeta-1}\right)b(p,\zeta).
\end{equation}
Here $b(p,\zeta)$ is an arbitrary function of its arguments that guarantees belonging of this kernel to the
space $\Ss'$. This representation is obviously preserved under evolutions defined by~(\ref{3}), so we get
\begin{align}\label{29:31}
 & B(p,\zeta;m,t,\q,\bz)=\delta\!\left(\bz-\dfrac{ip}{\zeta-1}\right)b(p,\zeta,m,t), \\
\intertext{where}
 &b(m,t,p,\zeta)=\exp\left(-ipt_1-t_2\dfrac{\zeta-2+\zeta^{-1}}{ip}\right)\zeta^{m}b(p,\zeta),\quad t\in\Rs^2,\quad
 m\in\Ns.
\end{align}
Here function $b(p,\zeta)$ is independent of $m$, $t_1$, and $t_2$. Below this function plays a role of the
scattering data. Correspondingly, we refer to $B$ as operator scattering data. Our next step is dressing
procedure that enables reconstruction of operator $\Lo$ in~(\ref{7}) for generic $\Uo$.

\section{Reconstruction of nonlinear integrable equation and Lax pair}

We introduce operator $\nu$ with kernel $\nu(p,\zeta;\bz)$, i.e., independent of the same variable $\q$ as
kernels of operators $A$ and $B$ as it was formulated in Condition~3. This operator is defined by
$\overline\partial$-problem with respect to variable $\bz$:
\begin{equation}
 \overline{\partial}_{\bz}\nu=\nu B,\label{26}
\end{equation}
where $\overline{\partial}_{\bz}$ is defined in~(\ref{2a13}). In order to determine operator $\nu$ uniquely
we normalize it to be unity operator at point of singularity of the kernel $L_{0}(p,\zeta;\q,\bz)$. Thanks
to~(\ref{L0}), (\ref{10a}) and~(\ref{T6})
\begin{equation}\label{L0expl}
  L_{0}(p,\zeta;\q,\bz)=(\bz-i\q)\delta(p)\delta_{c}(\zeta),
\end{equation}
so this kernel is entire function of $\bz$ linearly increasing at infinity. Thus we normalize operator $\nu$
in the following way:
\begin{equation}
 \nu(p,\zeta;\bz)=\delta(p)+O(\bz^{-1}),\quad \bz\to\infty.\label{2520}
\end{equation}
In what follows we assume the unique solvability of the problem~(\ref{26}), (\ref{2520}). Thanks to
analyticity of the kernel of $L_{0}(p,\zeta;\q,\bz)$, definition~(\ref{2a13}) and~(\ref{L0B}) we derive that
commutator $[L_0,\nu]$ obeys
\begin{equation}
 \overline{\partial}[L_0,\nu]=[L_0,\nu]B,\label{L0nu}
\end{equation}
i.e., the same Eq.~(\ref{26}). By~(\ref{L0F}) and~(\ref{2520}) kernel $[L_0,\nu](p;\bz)$ has some constant
(with respect to $\bz$) asymptotic behavior at infinity. Let us introduce operator $U$ with kernel
\begin{equation}\label{27}
  U(p,\zeta)=(\zeta-1)\lim_{\bz\to\infty}[\bz(\nu-I)(p,\zeta;\bz)],
\end{equation}
that by construction is independent of variables $\q$ and $\bz$. Then, thanks to~(\ref{L0F}), (\ref{26}) and
assuming unique solvability of Eq.~(\ref{26}) we get that $[L_0,\nu]=U\nu$, i.e.,
\begin{align}
 &L\nu=\nu L_{0},\label{23}\\
\intertext{where}
 &L=-iD+T-U,\label{Ldr}
\end{align}
that is extension (in the sense of~(\ref{1'}) or~(\ref{2aa})) of the dressed $\Lo$-operator~(\ref{7}). For
the kernel of operator $U$ in the $x$-space  we get by~(\ref{2b})
\begin{align}
 & U_{m,m'}(x,x')=u_{m}(x)\delta(x-x')\delta_{m,m'},\nonumber \\
 & u_{m}(x)=\int dp\oint\limits_{|\zeta|=1}\frac{d\zeta\,\zeta_{}^{-m-1}}{2\pi i}
 e_{}^{-ipx} U(p,\zeta),\label{Ux}
\end{align}
i.e., $U(x,x')$ is a diagonal matrix multiplication operator in correspondence to~(\ref{10:12}),
$U(x,x';q,h)=\Uo(x)\delta(x-x')$, where $\Uo(x)$ is potential of the $\Lo$-operator~(\ref{7}).
Eq.~(\ref{23}) means that $\nu$ is dressing (transformation) operator.

Dependence on ``times'' $m$, $t_1$, and $t_2$ of the dressing operator is switched on by means of the
corresponding dependence of the scattering data, i.e., instead of~(\ref{26}) we use
\begin{equation}
 \overline{\partial}_{\bz}\nu(m,t)=\nu(m,t) B(m,t),\label{26:1}
\end{equation}
where $B(m,t)$ is given by~(\ref{13:1})--(\ref{13:3}) and~(\ref{AT}). Normalization condition for $\nu(m,t)$
is preserved in the form~(\ref{2520}). By construction these flows commute and thanks to~(\ref{Bn1})
and~(\ref{Bn2}) we get
\begin{align}\label{num}
  &\nu(m+1,t)=T\nu(m,t)T^{-1},\\
  &\partial_{t_{1}}\nu(m,t)=-i[D,\nu(m,t)].\label{nut1}
\end{align}
In derivation of~(\ref{num}) we used that by~(\ref{T6}) and~(\ref{L0F})
\[(T{\nu}T^{-1})(p,\zeta;\bz)=\zeta\nu(p,\zeta;\bz),\]
so that
\begin{equation}\label{dbarT}
  \overline{\partial}_{\bz}(T\nu(m,t)T^{-1})=T(\overline{\partial}_{\bz}\nu(m,t))T^{-1}.
\end{equation}
The order $\bz^{-1}$ of equalities~(\ref{num}) and~(\ref{nut1}) gives thanks to~(\ref{27}) and~(\ref{Ux})
that
\begin{equation}\label{umt}
  u_{n}(x,t_1,t_2,m)=u_{n+m}(x+t_1,t_2),
\end{equation}
so we have to consider now dependence of this functions on $t_2$.

This can be performed in the same way, as above. Differentiating~(\ref{26}) with respect to $t_2$ we get
by~(\ref{13:3})
\begin{equation}\label{t2:1}
  \overline{\partial}\dfrac{\partial\nu(m,t)}{\partial{t_2}}=\nu(m,t) B(m,t)+\nu(m,t)[T^{-1},B(m,t)].
\end{equation}
Taking into account that kernel of $T^{-1}(p,\zeta;\bz)=\delta(p)\delta_{c}(\zeta)/\bz$ (see~(\ref{T6})), so
it has additional singularity, it is reasonable to multiply this equality by $T$ from the left. Then
by~(\ref{dbarT})
\begin{equation}\label{t2:2}
  \overline{\partial}(T\nu_{t_2}(m,t)+T\nu(m,t)T^{-1})=(T\nu_{t_2}(m,t)+T\nu(m,t)T^{-1})B.
\end{equation}
Thus we again get $\overline{\partial}$-equation~(\ref{26}) on operator $T\nu_{t_2}+T\nu T^{-1}$.
By~(\ref{2520}) limit of the kernel of this operator when $\bz$ exists and is finite, so we put
\begin{equation}\label{t2:3}
  C(p,\zeta)=\lim_{\bz\to\infty}(T\nu_{t_2}+T\nu T^{-1})(p,\zeta;\bz),
\end{equation}
so that
\begin{align}
  &T\nu_{t_2}+T\nu T^{-1}=C\nu,\nonumber \\
\intertext{that can be written as}
  & \partial_{t_2}\nu=M\nu-\nu T^{-1},\label{t2:4}\\
\intertext{where}
 &M=T^{-1}C.\label{t2:5}
\end{align}
It is easy to see that operator $M$ that appeared here is nothing but extension in the sense of
Sec.~\ref{resolvent} of the operator $\Mo$ in~(\ref{8}) of the Lax pair. Lax representation~(\ref{6}) here
is a simple consequence of the commutativity of all flows introduced by~(\ref{3}). Using now~(\ref{2520})
and~(\ref{27}) we get from the limit of~(\ref{t2:4}) for $\bz\to\infty$:
\begin{equation}\label{36}
  U_{t_2}(p,\zeta)=(1-\zeta)C(p,\zeta),
\end{equation}
that is extension of~(\ref{16:1}). Then~(\ref{16:2})follows from~(\ref{3}). Eq.~(\ref{10:12}) is definition
of $\phi_{n}(x)$ under condition that these functions and $\Uo(x)$ decay at space infinity.
Then~(\ref{10:13}) follows from~(\ref{16:2}) under assumption of the unity limit of matrix $\Co(x)$ at space
infinity. This completes reconstruction of the Lax pair and evolution equation~(\ref{5}) of the Toda lattice
by means of the representation~(\ref{3}) and the dressing procedure developed above.

\section{Concluding remarks}

Construction here was based on representation of elements $A$ and $B$ in~(\ref{1}) by means of the infinite
matrices. As the result we arrived to the infinite 2D Toda lattice. It is natural to expect that with choice
of $A$ as a finite permutation matrix one can derive Lax pairs and nonlinear equations corresponding to the
finite (periodic with respect to $n$) reductions of the infinite chain. Eq.~(\ref{3}) is close by its
meaning to condition of degeneracy of the dispersion in integrable models, see~\cite{Manakov}. But the
matrix generalizations of relations of the kind~(\ref{commut1}), (\ref{1}) and~(\ref{2}) considered
in~\cite{commut} show that approach based on commutator identities is more generic. It also enables to
introduce the dressing procedure presented here, that enables derivation of the Lax pair and integrable
equation corresponding to given commutator identity.

\end{document}